\documentclass[aps,prb,reprint,showkeys,showpacs]{revtex4-2}
\pdfoutput=1
\usepackage{microtype}
\usepackage{amsmath}
\usepackage{amssymb}
\usepackage{amsfonts}
\usepackage{graphicx}
\usepackage{bm}
\usepackage{siunitx}
\usepackage{color}
\usepackage{hyperref}
\hypersetup{%
    pdfborder = {0 0 0}
}

\allowdisplaybreaks
\usepackage{cleveref}
\usepackage[version=4]{mhchem}
\usepackage[T1]{fontenc}
\usepackage[utf8]{inputenc}
\usepackage[english]{babel}

\graphicspath{{./figures/}}

\renewcommand\vec{\bm} 
\newcommand{\uGrad}{\vec{\nabla}}

\newcommand{\ud}{\,{\mathrm{d}}}

\newcommand{\uC}{C} 
\newcommand{\uD}{D} 
\newcommand{\uE}{E} 
\newcommand{\uEE}{\mathcal{E}} 

\newcommand{\uHopf}{{\cal{H}}} 
\newcommand{\uh}{h} 
\newcommand{\uH}{H} 
\newcommand{\uvH}{\vec{H}}     
\newcommand{\uvHD}{\vec{H}_\mathrm{D}} 
\newcommand{\uK}{K}         

\newcommand{\uvM}{\vec{M}}
\newcommand{\uMi}{M_i}      
\newcommand{\uvm}{\vec{m}}

\newcommand{\udelta}{\delta}         

\newcommand{\uO}{O}         
\newcommand{\uP}{P_i}
\newcommand{\uPex}{P_\mathrm{EX}}
\newcommand{\uPa}{P_\mathrm{A}}
\newcommand{\uPz}{P_\mathrm{Z}}
\newcommand{\uPdm}{P_\mathrm{DM}}
\newcommand{\uPms}{P_\mathrm{MS}}
\newcommand{\udeltatrial}{\delta}
\newcommand{\up}{p_i}
\newcommand{\upex}{p_\mathrm{EX}}
\newcommand{\upa}{p_\mathrm{A}}
\newcommand{\upz}{p_\mathrm{Z}}
\newcommand{\updm}{p_\mathrm{DM}}
\newcommand{\upms}{p_\mathrm{MS}}
\newcommand{\uetaeq}{\eta_\mathrm{EQ}}

\newcommand{\uMs}{M_{\mathrm{S}}}
\newcommand{\uMssq}{M_{\mathrm{S}}^2}
\newcommand{\umuZ}{\mu_0}
\newcommand{\uq}{q} 
\newcommand{\uvs}{\vec{s}}  
\newcommand{\uX}{X}         
\newcommand{\uY}{Y}         
\newcommand{\uZ}{Z}         
\newcommand{\ur}{r}

\newcommand{\uvr}{\vec{r}}          

\begin{document}
\title{High-order concentric hopfions in a helimagnet}

\author{Konstantin L. Metlov}\email{metlov@donfti.ru}
\affiliation{Galkin Donetsk Institute for Physics and Engineering, R.~Luxembourg str.~72, Donetsk 283048, Russian Federation}

\author{Maksim M. Gordei}
\affiliation{Galkin Donetsk Institute for Physics and Engineering, R.~Luxembourg str.~72, Donetsk 283048, Russian Federation}

\date{\today}
\begin{abstract}
Magnetic hopfions are three-dimensional topological solitons: ``knots'' in the vector field of magnetization of magnetically ordered materials, which are a key element of the three-dimensional spintronics emerging today. In this work, a simple analytical model for a new type of magnetic hopfions --- ``concentric'' ones --- is constructed and solved. Their size and equilibrium energy are estimated, their stability region on the magnetic phase diagram of a helimagnet is mapped out, and the requirements for materials admitting equilibrium concentric hopfions with high Hopf index on a uniformly magnetized background are determined.
\end{abstract}


\maketitle

\section{Introduction}
Hopfions are three-dimensional topological solitons in vector fields~\cite{Faddeev1976}. They correspond to continuous, stable and spatially localized configurations of such fields and are characterized by a nonzero integer topological Hopf index $\uHopf$. If a preimage is defined as a spatial curve --- a compact set of points at which the field vector takes one and the same prescribed value --- then the Hopf index corresponds to the linking number of any two such curves associated with two distinct values of the field vector. Hopfions have been observed in the velocity field of superfluid helium flow~\cite{VM1977}, in liquid crystals~\cite{Chen2013,Ackerman2017}, in ferroelectrics~\cite{Lukyanchuk2025} and (which is the subject of the present work) in the vector fields of the local magnetization of magnets~\cite{KRRCVDCHSPFHSF2021}: $\uvM(\uvr)$, $|\uvM(\uvr)|=\uMs=const(\uvr)$, where $\uvr$ is a spatial coordinate. The topological aspects of magnetic hopfions are already understood quite well~\cite{DI79,BK11-book}, but the knowledge of their energetic aspects, which would allow one to determine the hopfion energy and their existence region in the parameter space of realistic helimagnets (stable hopfions cannot exist in ``classical'' ferromagnets with an inversion center~\cite{Hobart1963}), is still fragmentary.

Magnetic hopfions are interesting primarily from the fundamental point of view, since they are a physical realization of nontrivial vector-field topology, while their properties remain very poorly studied so far. But there is also a practical aspect. From the quaternionic representation of vector fields with hopfions, owing to the non-commutativity of quaternions, the non-Abelian nature of hopfion configurations follows as well~\cite{M2025quaternionic}. Simply put, if one takes a field configuration with two hopfions and smoothly interchanges them, the resulting field configuration turns out to be inequivalent to the original one. More complicated sequences of permutations are also possible, described by the braid algebra, on the basis of which, similarly to Majorana fermions, topological computations can be organized~\cite{Kitaev2003}.

However, the path towards the practical use of hopfions has only just begun. There are several problems. In most of the current experiments and numerical simulations, hopfions are artificially stabilized in finite nanoparticles with specially tailored highly anisotropic end faces. Although such configurations~\cite{KRRCVDCHSPFHSF2021} do possess a nonzero Hopf index, they are not suitable for realizing the braid algebra. Hopfions can be stable in bulk helimagnets~\cite{M2025}, but this imposes more stringent conditions on the material selection than in the case of hopfions in confined geometries. Besides hopfions, other magnetic phases also exist in a helimagnet, whose equilibrium regions overlap with the metastability region of hopfions~\cite{M2025}, which leads to the existence of such remarkable objects as heliknotons --- hopfions on a nonuniform helicoidal background --- which have also already been observed experimentally~\cite{Li2026}. The nonuniform background, however, hinders the hopfion motion required for realizing the braid algebra.

Hopfions on a uniformly magnetized background can be stable if this uniform magnetization is directed against the field~\cite{metlov2026tails}. Such a ``superheated'', yet still stable, uniform magnetization configuration can be created in a helimagnet with sufficiently strong uniaxial anisotropy, and hopfions locally lower its energy. This has been established for hopfions with unit Hopf index~\cite{metlov2026tails}. But is the existence of higher-order hopfions possible under such conditions?

Usually, such hopfions are described by the well-known Whitehead ansatz~\cite{whitehead1947} for a vector field with arbitrary Hopf index, written in bicomplex coordinates. Recently, however, a somewhat different expression for high-order hopfions has been obtained in Ref.~\cite{M2025quaternionic}. Homotopically, it is of course equivalent to the Whitehead ansatz, but it does not reduce to it directly. The vortex and antivortex filaments in such hopfions are circles with a common center; therefore, we will call these hopfions concentric. Their advantage is the high symmetry, which considerably simplifies the energy evaluation --- this is the main task of the present work.

\section{Concentric hopfions}
\begin{figure}[t]
    \begin{center}
        \includegraphics[width=\columnwidth]{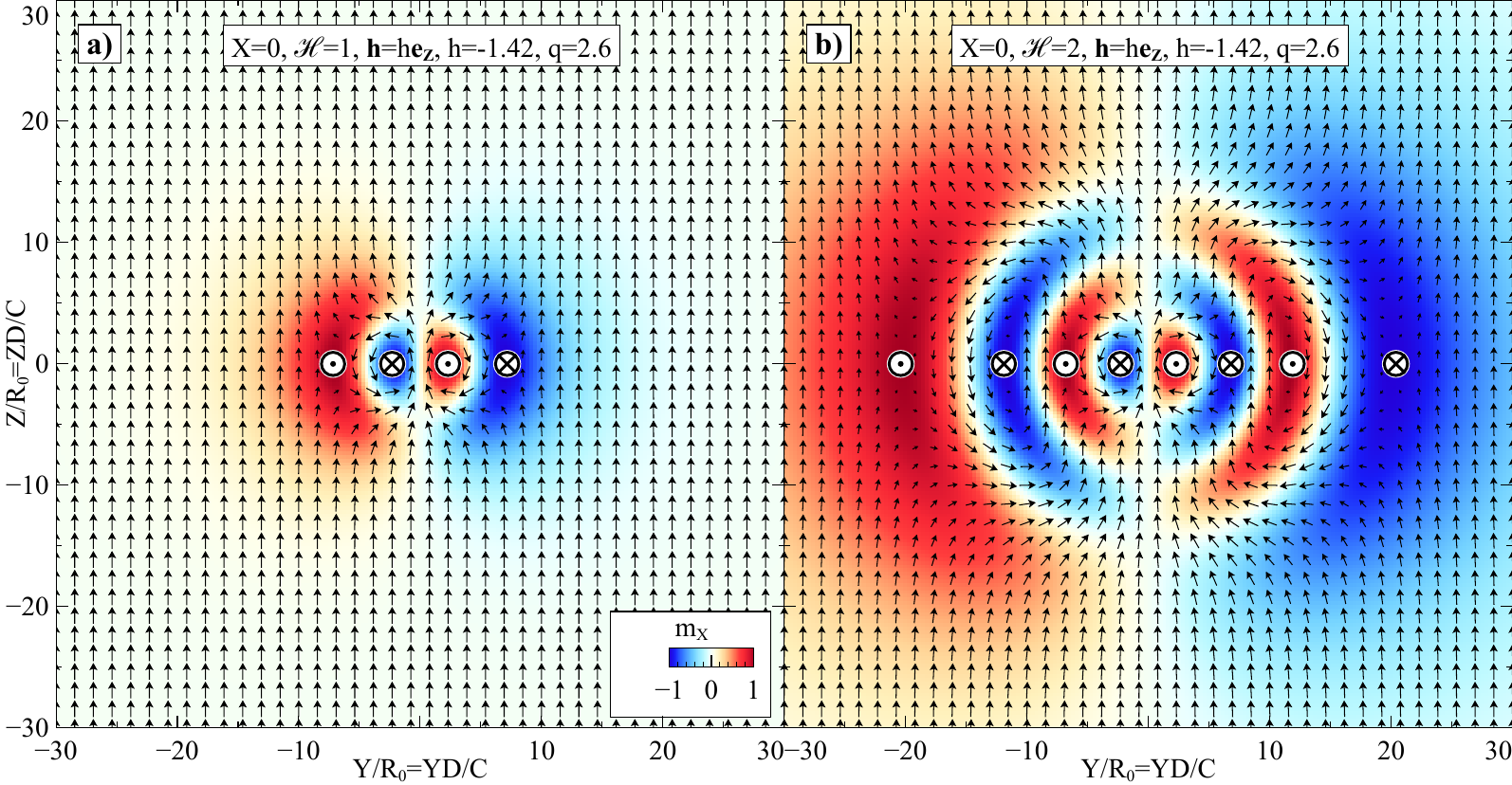}
    \end{center}
    \caption{\label{fig:hopfion}Cross section of the equilibrium hopfion~\eqref{eq:ansatz} with the profile~\eqref{eq:deltatrial} by the plane $X=0$ through its central axis $OZ$: a) with Hopf index $\uHopf=1$; b) with Hopf index $\uHopf=2$. The dimensionless magnetic field $h$ and the anisotropy parameter $q$ are defined after Eq.~\eqref{eq:F}.}
\end{figure}

According to the quaternionic ansatz~\cite{M2025quaternionic}, high-order hopfions correspond to magnetization distributions $\uvM(\uvr)$ containing $\uHopf$ hopfions with unit Hopf index whose centers coincide. These distributions take a particularly simple form in spherical coordinates:
\begin{align}
    \label{eq:ansatz}
    \uvm=&\left\{\cos\theta,-\sin\theta\sin\uHopf\delta(\ur),-\sin\theta\cos\uHopf\delta(\ur)\right\},
\end{align}
where the normalized magnetization vector $\uvm=\{m_{\rho}, m_{\phi}, m_{\theta}\}=\uvM/\uMs$, $|\uvm|=1$, is introduced, $\udelta(\ur)=4\arctan[e(\ur)/(1-e(\ur))]$, and $0\leq e(\ur)\leq1$, where $e(\ur)$ is the, so far unknown, profile function of the hopfion. These configurations are axially symmetric about the $\uO\uZ$ axis; examples of their cross sections are shown in Fig.~\ref{fig:hopfion}.

To find the equilibrium hopfion parameters, we write down the classical energy density of an infinite three-dimensional helimagnet containing the magnetization distribution $\uvM(\uvr)$:
\begin{equation}
\label{eq:energy}
\begin{aligned}
        \uE = & \frac{\uC}{2 \uMssq}\sum_{i=\uX,\uY,\uZ} \left|\uGrad\uMi\right|^2 +\frac{\uD}{\uMssq}\, \uvM\cdot[\uGrad\times\uvM] -\\
        &   -\umuZ \left(\uvM \cdot \uvH\right) - \frac{\umuZ}{2}\left(\uvM \cdot \uvHD\right)- \frac{\uK}{\uMssq}\left(\uvM \cdot \uvs\right)^2,
\end{aligned}
\end{equation}
where $C=2A$ is the exchange stiffness; $\uvr=\{\uX,\uY,\uZ\}$ is the radius vector of a point in the Cartesian coordinate system chosen such that at infinity $\lim_{|\uvr|\rightarrow\infty}\uvM = \{0,0,\uMs\}$; $D$ is the Dzyaloshinskii--Moriya interaction constant; $\umuZ$ is the vacuum magnetic permeability; $\uvH$ is the external magnetic field; $\uvHD$ is the demagnetizing field created by $\uvM(\uvr)$ according to Maxwell's equations; $\uK$ and $\uvs$ are the constant and the direction of the uniaxial anisotropy. We assume that the directions of $\uvH=\{0,0,\uH\}$ and $\uvs=\{0,0,1\}$ coincide and lie along the $\uO\uZ$ axis.

Substituting~\eqref{eq:ansatz} into the energy density~\eqref{eq:energy} and integrating over the whole space in spherical coordinates, the angular integrals can be taken analytically, and for the normalized total energy we obtain $\uEE=(D/C^2)\int_V E \ud^3\uvr = \int_0^\infty F \ud \rho$, where
\begin{equation}
    \label{eq:F}
    \begin{split}
        &F = \eta \uPex+ \eta^2 \uPdm + \eta^3 \left( h \uPz + \frac{q}{2} \uPa + \mu^2 \uPms \right),
    \end{split}
\end{equation}
and $\eta = D R/C$ is the characteristic spatial scale (radius) of the hopfion normalized by $R_0 = C/D$, which appears upon passing to the dimensionless spatial coordinate $\rho=\ur/R$; $h=\umuZ \uMs C H/D^2$ and $q=2CK/D^2$ are the normalized magnetic field and uniaxial anisotropy constant; the dimensionless parameter $\mu^2 = \umuZ\uMssq C/D^2$, also known as the conical phase susceptibility~\cite{GWG2017}, characterizes the strength of the magnetostatic interaction. The radial energy densities then take the following form:
\begin{subequations}
    \label{eq:Pi}
    \begin{align}
        \uPex &= \frac{4}{3} \pi \left(4 - 4 \cos\uHopf\delta + \rho^2 \uHopf^2\delta^{\prime2}\right), \label{eq:Pex}\\
        \uPdm &= -\frac{8}{3} \pi \rho \left(2 \sin\uHopf\delta + \rho \uHopf\delta^\prime\right), \label{eq:Pdm}\\
        \uPz  &= \frac{8}{3} \pi \rho^2 \left(1 - \cos\uHopf\delta\right), \label{eq:Pz}\\
        \uPa  &= \frac{16}{15} \pi \rho^2 \left(2 - \cos\uHopf\delta - \cos2\uHopf\delta\right), \label{eq:Pa}\\
        \uPms &= \frac{8 \pi \left(1 - \cos\uHopf\delta\right)}{9 \rho} \int_0^\rho \tau^2 \left[1 - \cos\uHopf\delta(\tau)\right] d\tau, \label{eq:Pms}
    \end{align}
\end{subequations}
where $\delta\equiv\delta(\rho)$ and $\delta^\prime\equiv\ud\delta(\rho)/\ud\rho$.

\begin{figure} [t]
    \begin{center}
        \includegraphics[width=\columnwidth]{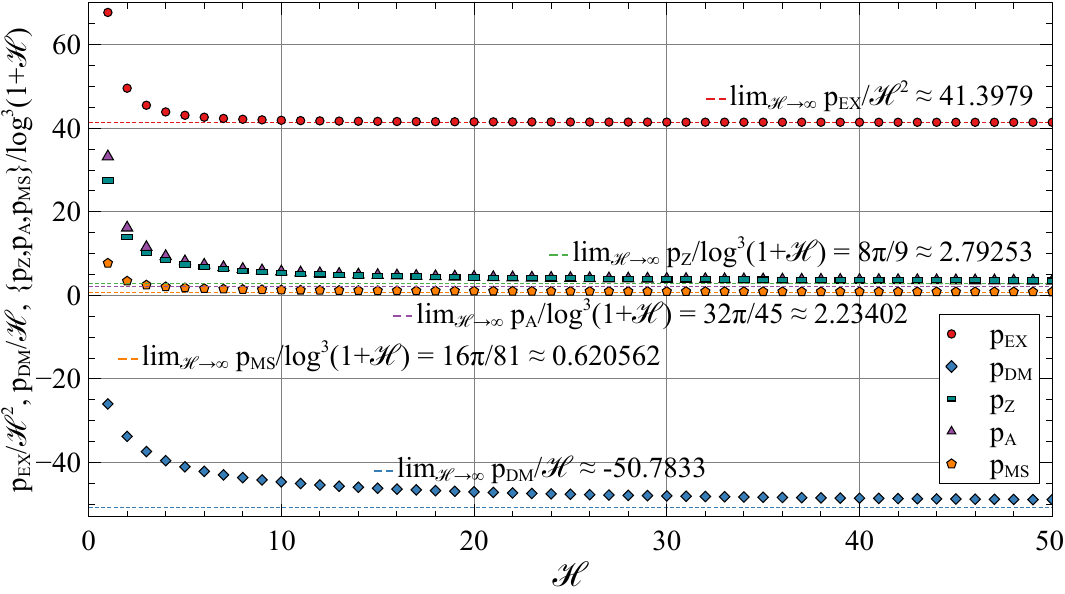}
    \end{center}
    \caption{\label{fig:enfunc}Normalized energy parameters $\up$ and their asymptotics at large Hopf indices $\uHopf$.}
\end{figure}

The equilibrium hopfion profile $\delta(\ur)$ is determined by the minimum of the functional $\uEE$ with the radial energy density~\eqref{eq:F}. Even in this one-dimensional formulation, finding this minimum is a difficult task. However, one can construct a simple analytical ansatz for the profile function~$\udeltatrial(\ur)$, smoothly matching the exact asymptotics of this function at small and large $\ur$. For the high-order hopfions considered here, at $\mu=0$ these asymptotics (when expressed in terms of $e(\ur)$) turn out to be identical to the asymptotics of the $\uHopf=1$ hopfions obtained in Ref.~\cite{metlov2026tails}. For $\rho\gg1$ this is not surprising, since the asymptotic behavior proportional to the modified spherical Bessel function is universal for spatially localized magnetization distributions in three dimensions with the micromagnetic Hamiltonian~\eqref{eq:energy}, provided the magnetostatic energy can be neglected. For $\rho\ll1$, the coincidence of the asymptotics is caused by the fact that the internal structure of a concentric hopfion is identical to that of a unit hopfion up to a scale factor. Smoothly matching these two asymptotics, we obtain an ansatz for the trial profile function, asymptotically exact for both small and large $\rho$:
\begin{equation}
 \label{eq:deltatrial}
 \udeltatrial(\rho) = 4 \arctan\left(\frac{2 e^{\rho } \left(\rho ^3+1\right)}{2 \rho ^2-\sqrt{3} \rho
   +\rho +2}-1\right).
\end{equation}
This expression contains no parameters in the dimensionless variables. After substituting~\eqref{eq:deltatrial} into~\eqref{eq:Pi}, the integrals $\up=\int_0^\infty\uP\ud \rho$ depend only on the Hopf index $\uHopf$; their values for $\uHopf\leq10$ are given in Table~\ref{tab:params}.
\begin{table}[t]
    \caption{\label{tab:params}Normalized energy parameters $\up$ for some small values of the Hopf index. The values in the first row coincide with those obtained in Ref.~\cite{metlov2026tails}.}
    \begin{ruledtabular}
\begin{tabular}{rrrrrr}
    \hline
    $\uHopf$ & $\upex$ & $\updm$ & $\upz$ & $\upa$  & $\upms$\\
    1 & 67.7287 & -25.984 & 9.163 & 11.0663 & 2.56789\\
    2 & 198.363 & -67.5078 & 18.5028 & 21.4669 & 4.53744\\
    3 & 409.583 & -112.176 & 27.1755 & 30.703 & 6.61331\\
    4 & 702.64 & -158.234 & 35.1645 & 39.0365 & 8.54786\\
    5 & 1077.84 & -205.183 & 42.6023 & 46.6766 & 10.3511\\
    6 & 1535.41 & -252.739 & 49.582 & 53.7618 & 12.0436\\
    7 & 2075.47 & -300.733 & 56.1722 & 60.3894 & 13.6403\\
    8 & 2698.09 & -349.06 & 62.4266 & 66.6311 & 15.1534\\
    9 & 3403.33 & -397.647 & 68.3875 & 72.5418 & 16.5933\\
    10 & 4191.22 & -446.444 & 74.0892 & 78.1645 &  17.9683\\
    \hline
\end{tabular}
\end{ruledtabular}
\end{table}
The asymptotic behavior of these quantities at large $\uHopf$ is shown in Fig.~\ref{fig:enfunc}.

The total energy $\uEE$ now depends only on the spatial scale --- the hopfion radius $\eta$
\begin{equation}
  \label{eq:energyTrial}
  \uEE  = \eta \upex + \eta^2 \updm + \eta^3 \left(h \upz + \frac{q}{2} \upa + \mu^2 \upms\right),
\end{equation}
where $\up$ are just numbers. The minimization of the energy with respect to $\eta$ reduces to solving a quadratic equation, one of whose roots
\begin{align*}
\uetaeq = & \frac{d-2 \updm}{6 \uh \upz+3 \uq \upa+6 \mu ^2 \upms},\\
d = & \sqrt{4 \updm^2-6 \upex \left(2 \uh \upz+\uq \upa+2 \mu ^2 \upms\right)}
\end{align*}
corresponds to the energy minimum, while the other one --- to the maximum.
The equilibrium hopfion radius grows almost linearly with increasing index $\uHopf$, which is also clearly seen in Fig.~\ref{fig:hopfion}. This means that the inner vortex and antivortex rings have a certain equilibrium size, determined by the material properties, towards which they would tend in the absence of pressure from the outer rings. Therefore, adding rings (increasing the Hopf index) proportionally enlarges the hopfion itself.

\begin{figure}[t]
    \begin{center}
        \includegraphics[width=\columnwidth]{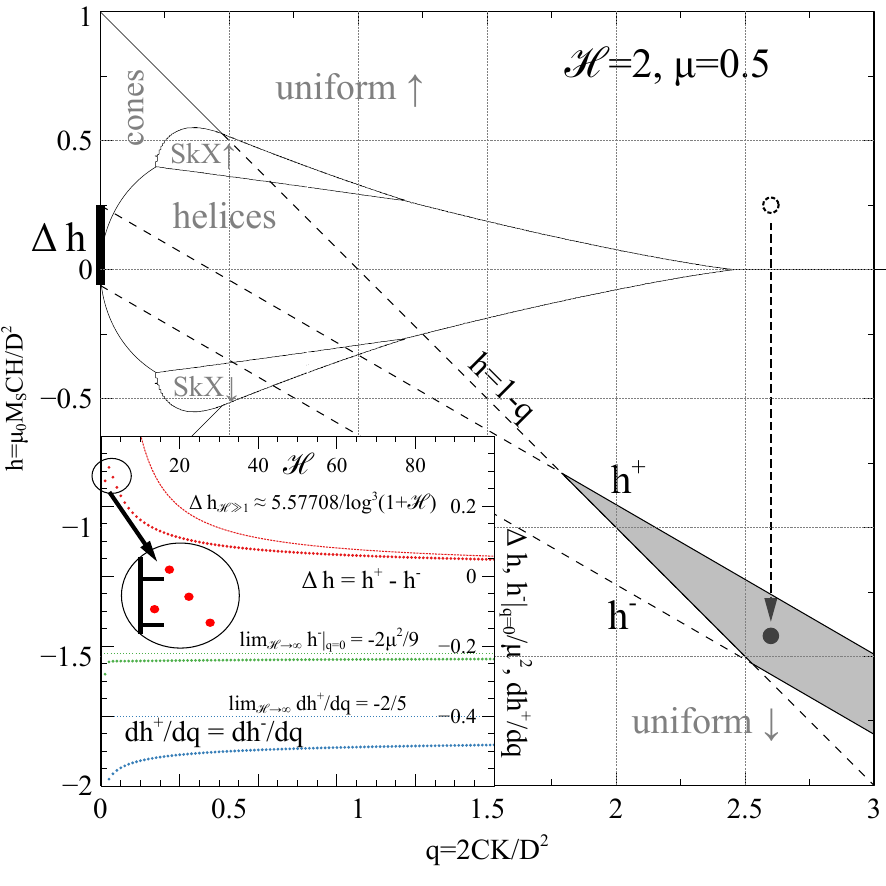}
    \end{center}
    \caption{\label{fig:phasediagram}(left) Magnetic phase diagram with the stability region of hopfions with Hopf index $\uHopf=2$ overlaid; the filled circle marks the set of hopfion parameters shown in Fig.~\ref{fig:hopfion}; the open circle shows a possible initial equilibrium state magnetized in the positive direction of the $OZ$ axis, which, by applying a negative field, is turned into a metastable uniform background for the hopfions; the inset shows the dependences of the parameters of the lines outlining the stability region on the Hopf index $\uHopf$ and the asymptotics of these dependences.}
\end{figure}

Hopfions are energetically favorable when their energy is lower than that of the uniform background. The corresponding region in the $\uq$--$\uh$ parameter space is bounded by two straight lines
\begin{equation*}
 h^{-} = \frac{-q \upa-2 \mu ^2 \upms}{2 \upz},\quad
 h^{+} = h^{-} + \frac{\updm^2}{4\upex \upz}.
\end{equation*}
The vertical distance between them equals $\Delta h = \updm^2/(4\upex \upz)$, and their slope is $\ud \uh^{+}/\ud q = \ud \uh^{-}/\ud q = - \upa/(2\upz)$. In this simple model, the role of the magnetostatic interaction reduces to shifting the stability region vertically downward by the amount $\mu^2\upms/\upz$. In addition, the stability region of hopfions on the uniform background is limited by the stability condition of the background itself, $h>1-q$, corresponding to its instability towards a transition into the conical phase followed by a transition into the uniform state magnetized along the field. As $\mu$ increases, this condition shifts the hopfion stability region towards larger anisotropy $\uq$ and larger values of $|\uh|$. Therefore, in general, the magnetostatic interaction destabilizes hopfions at given $\uq$ and $\uh$. Note, however, that the simple model considered here is based on a trial function that does not take the magnetostatic interaction into account and is therefore, strictly speaking, valid only for small $\mu^2$. An example of the stability region for hopfions with $\uHopf=2$ is shown in Fig.~\ref{fig:phasediagram}.

Note that Ref.~\cite{metlov2026tails} took into account only the stability of the uniform phase magnetized against the field with respect to zero-wave-vector perturbations (uniform rotation), determined by the condition $h>-q$. In reality, in a helimagnet, perturbations with a finite wave vector, corresponding to the transition into the conical phase, turn out to be more energetically favorable. This does not change the principal conclusions of Ref.~\cite{metlov2026tails}, but shifts the hopfion stability region towards larger values of $\uq$, as shown in Fig.~\ref{fig:phasediagram}.

The same figure also shows the dependences on the Hopf index, and their asymptotics, of the three parameters fully determining the stability region: $\Delta h$, $\ud h^{+}/\ud q$ and $h^{-}|_{q=0}$. In general, with increasing topological index, the hopfion stability region shrinks, since $\Delta h$ tends to zero as $\uHopf\rightarrow\infty$. However, it decreases so slowly ($\propto 1/\log^3(1+\uHopf)$) that, in practice, the realization of concentric hopfions with arbitrarily large Hopf index is possible. This dependence has an interesting feature as well: hopfions with $\uHopf=2$ and $\uHopf=3$ have a wider stability region than hopfions with $\uHopf=1$. Owing to this, ``sifting'' of hopfions by topological index becomes possible by varying the magnetic field strength.

The stability region also tilts slightly with increasing Hopf index, and this tilt approaches a finite limit as $\uHopf\rightarrow\infty$, whereas its downward shift with increasing $\mu^2$ due to the magnetostatic interaction depends very weakly on $\uHopf$.

\section{Conclusions}
In this work, a simple analytical model for a new type of magnetic hopfions in a helimagnet has been constructed and solved. The equilibrium profiles have been obtained, and the hopfion stability region has been mapped out in the parameter space: magnetic field -- anisotropy, taking the magnetostatic interaction into account.

The analysis shows that concentric hopfions of practically arbitrarily high order can exist in a helimagnet on top of a uniform state magnetized against the applied magnetic field. The range of material parameters capable of hosting such hopfions has also been determined. The material should be an anisotropic helimagnet with low saturation magnetization (since magnetostatic effects destabilize hopfions) and a relatively small Dzyaloshinskii--Moriya constant.

None of the helimagnets widely studied today in the search for topological states possesses these properties; therefore, the hopfions observed in experiments either require additional stabilization in confined geometries, or exist on top of nonuniform magnetization states, which hinders their free motion. Moreover, most materials with the required anisotropy values will not exhibit, in equilibrium, the chiral phases typical of helimagnets because the Dzyaloshinskii--Moriya interaction is not strong enough for that, yet they will be able to support metastable hopfions.

Nevertheless, the current progress in materials science, including the use of neural-network technologies~\cite{Lyakhov2013}, gives hope that materials satisfying these criteria can be found or engineered. This, in turn, would open the way to three-dimensional spintronics, in which hopfions, owing to their non-Abelian nature, could become a key element for massively parallel topological computations.

This work was supported by the Russian Science Foundation, project no. RSF 25-22-00076.

%

\end{document}


\title{Концентрические хопфионы высоких порядков в гелимагнетике}

\author{К.Л. Метлов}\email{metlov@donfti.ru}
\affiliation{Донецкий физико-технический институт им. А.А. Галкина, ул. Р. Люксембург 72, Донецк, ДНР, Россия 283048}

\author{М.М. Гордей}
\affiliation{Донецкий физико-технический институт им. А.А. Галкина, ул. Р. Люксембург 72, Донецк, ДНР, Россия 283048}

\date{\today}
\begin{abstract}
Магнитные хопфионы --- трёхмерные топологические солитоны: ``узлы'' в векторном поле намагниченности магнитоупорядоченных материалов, являющиеся ключевым элементом зарождающейся сегодня трёхмерной спинтроники. В работе построена и решена простая аналитическая модель для нового типа магнитных хопфионов --- ``концентрических''. Проведена оценка их размера и равновесной энергии, построена область их стабильности на магнитной фазовой диаграмме гелимагнетика, определены требования к материалам, допускающим существование в них равновесных концентрических хопфионов с высоким индексом Хопфа на фоне однородной намагниченности.
\end{abstract}


\maketitle

\section{Введение}
Хопфионы --- трёхмерные топологические солитоны в векторных полях~\cite{Faddeev1976}. Они соответствуют непрерывным, стабильным и пространственно локализованным конфигурациям этих полей и характеризуются ненулевым целочисленным топологическим индексом Хопфа $\uHopf$. Если определить прообраз как пространственную кривую --- компактное множество точек, в которых вектор поля принимает одно и то же наперёд заданное значение, --- то индекс Хопфа соответствует коэффициенту зацепления любых двух таких кривых, отвечающих двум различным значениям векторного поля. Хопфионы наблюдаются в векторном поле скоростей течения сверхтекучего гелия~\cite{VM1977}, в жидких кристаллах~\cite{Chen2013,Ackerman2017}, сегнетоэлектриках~\cite{Lukyanchuk2025} и (что является предметом настоящей работы) в векторных полях локальной намагниченности магнетиков~\cite{KRRCVDCHSPFHSF2021}: $\uvM(\uvr)$, $|\uvM(\uvr)|=\uMs=const(\uvr)$, где $\uvr$ --- пространственная координата. Топологические аспекты магнитных хопфионов поняты уже достаточно хорошо~\cite{DI79,BK11-book}, но вот знания об их энергетических аспектах, позволяющие определить энергию хопфионов и область их существования в пространстве параметров реалистичных гелимагнетиков (в ``классических'' ферромагнетиках с центром инверсии стабильных хопфионов быть не может~\cite{Hobart1963}) --- пока ещё отрывочны.

Магнитные хопфионы интересны прежде всего с фундаментальной точки зрения, поскольку являются физической реализацией нетривиальной топологии векторного поля, а их свойства пока очень слабо изучены. Но есть и практический аспект. Из кватернионного представления векторных полей с хопфионами, в силу некоммутативности кватернионов, следует также неабелевость хопфионных конфигураций~\cite{M2025quaternionic}. Проще говоря, если взять конфигурацию поля с двумя хопфионами и плавно поменять эти хопфионы местами, конфигурация поля получится неэквивалентной исходной. Возможны и более сложные последовательности перестановок, описываемые алгеброй кос, на основе которых, подобно фермионам Майораны, можно организовать топологические вычисления~\cite{Kitaev2003}.

Но путь к практическому использованию хопфионов пока ещё только начинается. Есть несколько проблем. В большинстве современных экспериментов и численных симуляций хопфионы искусственно стабилизируются в конечных наночастицах со специально подобранными высокоанизотропными торцами. Хотя такие конфигурации~\cite{KRRCVDCHSPFHSF2021} действительно обладают ненулевым индексом Хопфа, для реализации алгебры кос они не подходят. Хопфионы могут быть стабильными в объёмном гелимагнетике~\cite{M2025}, но это налагает более жёсткие условия на отбор материалов, чем в случае хопфионов в ограниченной геометрии. Кроме хопфионов, в гелимагнетике существуют и другие магнитные фазы, область равновесия которых накладывается на область метастабильности хопфионов~\cite{M2025}, что приводит к существованию таких замечательных объектов, как геликнотоны --- хопфионы на фоне неоднородной геликоидальной структуры, которые тоже уже наблюдались экспериментально~\cite{Li2026}. Однако неоднородный фон затрудняет перемещение хопфионов, необходимое для реализации алгебры кос.

Хопфионы на фоне однородной намагниченности могут быть стабильными в случае, если эта однородная намагниченность направлена против поля~\cite{metlov2026tails}. Такая ``перегретая'', но всё же стабильная, однородная конфигурация намагниченности может быть создана в гелимагнетике с достаточно сильной одноосной анизотропией, и хопфионы локально понижают её энергию. Это установлено для хопфионов с единичным индексом Хопфа~\cite{metlov2026tails}. Но возможно ли существование в таких условиях хопфионов более высокого порядка?

Обычно для описания таких хопфионов используется хорошо известный анзац Уайтхеда~\cite{whitehead1947} для векторного поля с произвольным индексом Хопфа, записанный в бикомплексных координатах. Однако недавно в работе~\cite{M2025quaternionic} было получено несколько другое выражение для хопфионов высоких порядков. Гомотопически оно конечно же эквивалентно анзацу Уайтхеда, но напрямую к нему не сводится. Вихревые и антивихревые нити в таких хопфионах являются окружностями с общим центром, поэтому мы будем называть эти хопфионы концентрическими. Их достоинством является высокая симметрия, значительно упрощающая оценку энергии, что и стало основной задачей данной работы.

\section{Концентрические хопфионы}
\begin{figure}[t]
    \begin{center}
        \includegraphics[width=\columnwidth]{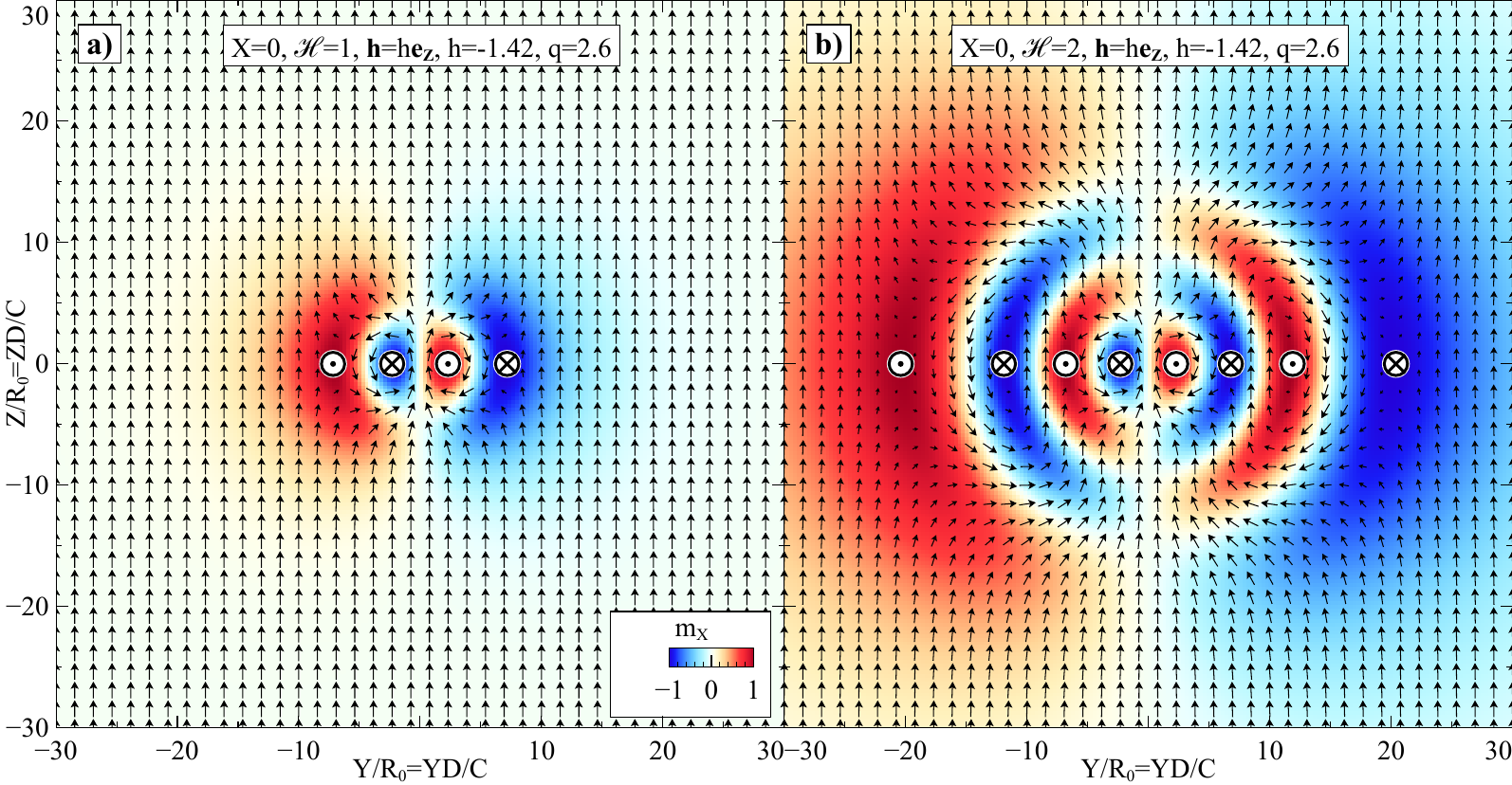}
    \end{center}
    \caption{\label{fig:hopfion}Поперечное сечение равновесного хопфиона~\eqref{eq:ansatz} с профилем~\eqref{eq:deltatrial} плоскостью $X=0$ через его центральную ось $OZ$: а) с индексом Хопфа $\uHopf=1$; б) с индексом Хопфа $\uHopf=2$. Безразмерное магнитное поле $h$ и параметр анизотропии $q$ определены после формулы~\eqref{eq:F}.}
\end{figure}

Согласно кватернионному анзацу~\cite{M2025quaternionic}, хопфионы высоких порядков соответствуют распределениям намагниченности $\uvM(\uvr)$, содержащим $\uHopf$ хопфионов с единичным индексом Хопфа, центры которых совпадают. Эти распределения принимают особенно простой вид в сферической системе координат:
\begin{align}
    \label{eq:ansatz}
    \uvm=&\left\{\cos\theta,-\sin\theta\sin\uHopf\delta(\ur),-\sin\theta\cos\uHopf\delta(\ur)\right\},
\end{align}
где введён нормированный вектор намагниченности $\uvm=\{m_{\rho}, m_{\phi}, m_{\theta}\}=\uvM/\uMs$, $|\uvm|=1$, $\udelta(\ur)=4\arctan[e(\ur)/(1-e(\ur))]$, а $0\leq e(\ur)\leq1$, где $e(\ur)$ --- пока неизвестная функция профиля хопфиона. Эти конфигурации аксиально-симметричны вокруг оси $\uO\uZ$, примеры их поперечного сечения приведены на рис.~\ref{fig:hopfion}.

Для нахождения равновесных параметров хопфиона запишем классическую плотность энергии бесконечного трёхмерного гелимагнетика, содержащего распределение намагниченности $\uvM(\uvr)$:
\begin{equation}
\label{eq:energy}
\begin{aligned}
        \uE = & \frac{\uC}{2 \uMssq}\sum_{i=\uX,\uY,\uZ} \left|\uGrad\uMi\right|^2 +\frac{\uD}{\uMssq}\, \uvM\cdot[\uGrad\times\uvM] -\\
        &   -\umuZ \left(\uvM \cdot \uvH\right) - \frac{\umuZ}{2}\left(\uvM \cdot \uvHD\right)- \frac{\uK}{\uMssq}\left(\uvM \cdot \uvs\right)^2,
\end{aligned}
\end{equation}
где $C=2A$ — обменная жёсткость; $\uvr=\{\uX,\uY,\uZ\}$ — радиус-вектор точки в декартовой системе координат, выбранной так, что на бесконечности $\lim_{|\uvr|\rightarrow\infty}\uvM = \{0,0,\uMs\}$; $D$ — величина взаимодействия Дзялошинского–Мория; $\umuZ$ — магнитная проницаемость вакуума; $\uvH$ — внешнее магнитное поле; $\uvHD$ — размагничивающее поле, создаваемое $\uvM(\uvr)$ согласно уравнениям Максвелла; $\uK$ и $\uvs$ — константа и направление одноосной анизотропии. Мы предполагаем, что направления $\uvH=\{0,0,\uH\}$ и $\uvs=\{0,0,1\}$ совпадают и направлены вдоль оси $\uO\uZ$.

Подставим~\eqref{eq:ansatz} в плотность энергии~\eqref{eq:energy} и проинтегрируем по всему пространству в сферической системе координат. Угловые интегралы тогда можно взять аналитически, а для нормированной полной энергии получим $\uEE=(D/C^2)\int_V E \ud^3\uvr = \int_0^\infty F \ud \rho$, где
\begin{equation}
    \label{eq:F}
    \begin{split}
        &F = \eta \uPex+ \eta^2 \uPdm + \eta^3 \left( h \uPz + \frac{q}{2} \uPa + \mu^2 \uPms \right),
    \end{split}
\end{equation}
а $\eta = D R/C$ --- нормированный в единицах $R_0 = C/D$ характерный пространственный масштаб (радиус) хопфиона, возникающий при переходе к безразмерной пространственной координате $\rho=\ur/R$; $h=\umuZ \uMs C H/D^2$ и $q=2CK/D^2$ --- нормированные магнитное поле и константа одноосной анизотропии; безразмерный параметр $\mu^2 = \umuZ\uMssq C/D^2$, известный также как восприимчивость конической фазы~\cite{GWG2017}, характеризует силу магнитостатического взаимодействия. Радиальные плотности энергии при этом имеют следующий вид:
\begin{subequations}
    \label{eq:Pi}
    \begin{align}
        \uPex &= \frac{4}{3} \pi \left(4 - 4 \cos\uHopf\delta + \rho^2 \uHopf^2\delta^{\prime2}\right), \label{eq:Pex}\\
        \uPdm &= -\frac{8}{3} \pi \rho \left(2 \sin\uHopf\delta + \rho \uHopf\delta^\prime\right), \label{eq:Pdm}\\
        \uPz  &= \frac{8}{3} \pi \rho^2 \left(1 - \cos\uHopf\delta\right), \label{eq:Pz}\\
        \uPa  &= \frac{16}{15} \pi \rho^2 \left(2 - \cos\uHopf\delta - \cos2\uHopf\delta\right), \label{eq:Pa}\\
        \uPms &= \frac{8 \pi \left(1 - \cos\uHopf\delta\right)}{9 \rho} \int_0^\rho \tau^2 \left[1 - \cos\uHopf\delta(\tau)\right] d\tau, \label{eq:Pms}
    \end{align}
\end{subequations}
где $\delta\equiv\delta(\rho)$ и $\delta^\prime\equiv\ud\delta(\rho)/\ud\rho$.

\begin{figure} [t]
    \begin{center}
        \includegraphics[width=\columnwidth]{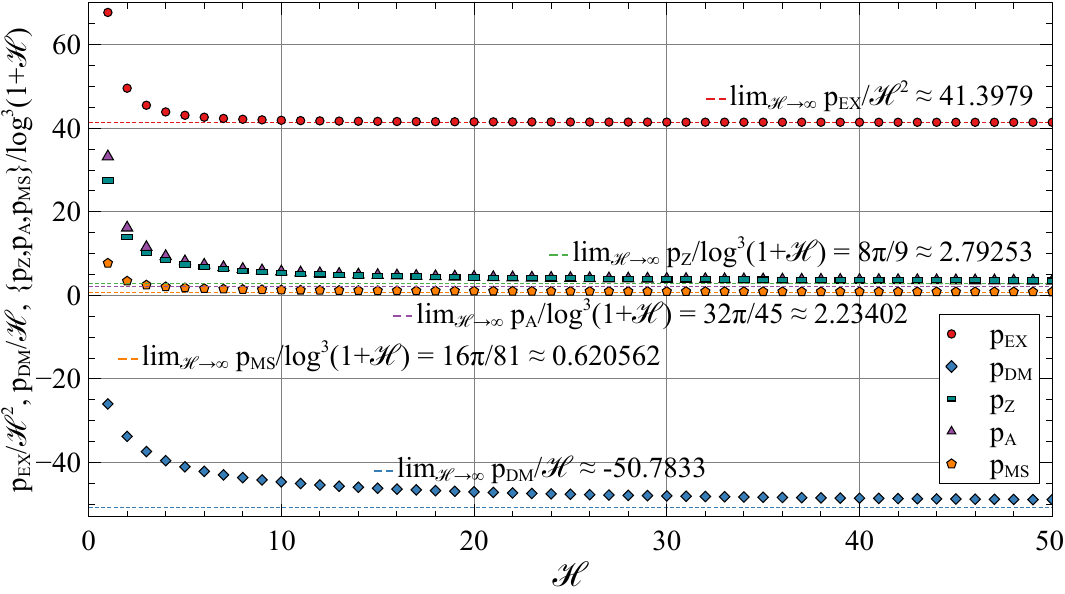}
    \end{center}
    \caption{\label{fig:enfunc}Нормированные энергетические параметры $\up$ и их асимптотики при больших индексах Хопфа $\uHopf$.}
\end{figure}

Равновесный профиль хопфиона $\delta(\ur)$ определяется минимумом функционала $\uEE$ с радиальной плотностью энергии~\eqref{eq:F}. Даже в такой одномерной постановке нахождение этого минимума --- сложная задача. Однако можно построить простой аналитический анзац для функции профиля~$\udeltatrial(\ur)$, сшивающий точные асимптотики этой функции при малых и больших значениях $\ur$. Для рассматриваемых здесь хопфионов высокого порядка в случае $\mu=0$ эти асимптотики (если их представить в терминах $e(\ur)$) получаются идентичными асимптотикам хопфионов с $\uHopf=1$, полученным в работе~\cite{metlov2026tails}. В случае $\rho\gg1$ это и неудивительно, поскольку асимптотическое поведение, пропорциональное модифицированной сферической функции Бесселя, является универсальным для пространственно локализованных распределений намагниченности в трёх измерениях с микромагнитным гамильтонианом~\eqref{eq:energy}, если магнитостатической энергией можно пренебречь. В случае $\rho\ll1$ совпадение асимптотик вызвано тем, что внутренняя структура концентрического хопфиона идентична структуре единичного хопфиона с точностью до масштаба. Гладко сшивая эти две асимптотики, получим анзац для пробной функции профиля, асимптотически точный для малых и для больших $\rho$:
\begin{equation}
 \label{eq:deltatrial}
 \udeltatrial(\rho) = 4 \arctan\left(\frac{2 e^{\rho } \left(\rho ^3+1\right)}{2 \rho ^2-\sqrt{3} \rho
   +\rho +2}-1\right).
\end{equation}
Это выражение в безразмерных переменных не содержит параметров. После подстановки~\eqref{eq:deltatrial} в~\eqref{eq:Pi} интегралы $\up=\int_0^\infty\uP\ud \rho$ зависят только от индекса Хопфа $\uHopf$; их значения для $\uHopf\leq10$ приведены в таблице~\ref{tab:params}.
\begin{table}[t]
    \caption{\label{tab:params}Нормированные энергетические параметры $\up$ при некоторых малых значениях индекса Хопфа. Значения в первой строке совпадают с полученными в работе~\cite{metlov2026tails}.}
    \begin{ruledtabular}
\begin{tabular}{rrrrrr}
    \hline
    $\uHopf$ & $\upex$ & $\updm$ & $\upz$ & $\upa$  & $\upms$\\
    1 & 67.7287 & -25.984 & 9.163 & 11.0663 & 2.56789\\
    2 & 198.363 & -67.5078 & 18.5028 & 21.4669 & 4.53744\\
    3 & 409.583 & -112.176 & 27.1755 & 30.703 & 6.61331\\
    4 & 702.64 & -158.234 & 35.1645 & 39.0365 & 8.54786\\
    5 & 1077.84 & -205.183 & 42.6023 & 46.6766 & 10.3511\\
    6 & 1535.41 & -252.739 & 49.582 & 53.7618 & 12.0436\\
    7 & 2075.47 & -300.733 & 56.1722 & 60.3894 & 13.6403\\
    8 & 2698.09 & -349.06 & 62.4266 & 66.6311 & 15.1534\\
    9 & 3403.33 & -397.647 & 68.3875 & 72.5418 & 16.5933\\
    10 & 4191.22 & -446.444 & 74.0892 & 78.1645 &  17.9683\\
    \hline
\end{tabular}
\end{ruledtabular}
\end{table}
Асимптотическое поведение этих величин при больших $\uHopf$ изображено на рис.~\ref{fig:enfunc}.

Полная энергия $\uEE$ теперь зависит только от пространственного масштаба --- радиуса хопфиона $\eta$
\begin{equation}
  \label{eq:energyTrial}
  \uEE  = \eta \upex + \eta^2 \updm + \eta^3 \left(h \upz + \frac{q}{2} \upa + \mu^2 \upms\right),
\end{equation}
где $\up$ --- просто числа. Минимизация энергии по $\eta$ сводится к решению квадратного уравнения, один из корней которого
\begin{align*}
\uetaeq = & \frac{d-2 \updm}{6 \uh \upz+3 \uq \upa+6 \mu ^2 \upms},\\
d = & \sqrt{4 \updm^2-6 \upex \left(2 \uh \upz+\uq \upa+2 \mu ^2 \upms\right)}
\end{align*}
соответствует минимуму энергии, а второй --- максимуму.
Равновесный радиус хопфиона увеличивается почти линейно с ростом индекса $\uHopf$, что также хорошо видно из рис.~\ref{fig:hopfion}. Это значит, что у внутренних вихревых и антивихревых колец есть некоторый равновесный размер, определяемый свойствами материала, к которому они стремились бы в отсутствие давления со стороны внешних колец. Поэтому добавление колец (увеличение индекса Хопфа) пропорционально увеличивает и сам хопфион.

\begin{figure}[t]
    \begin{center}
        \includegraphics[width=\columnwidth]{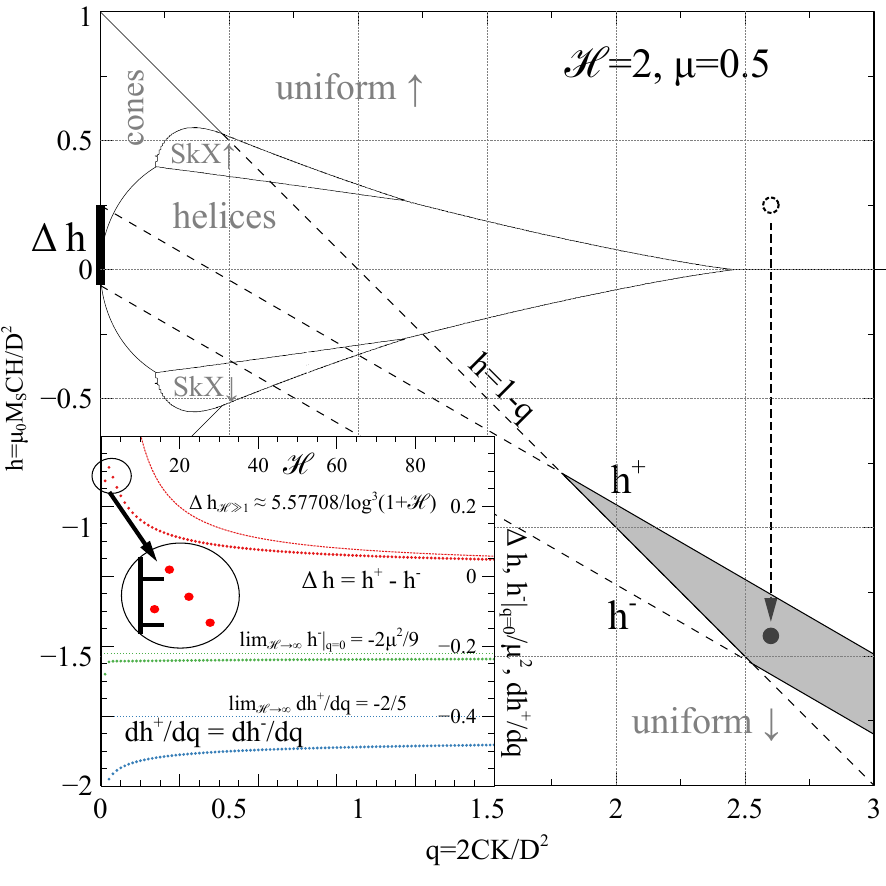}
    \end{center}
    \caption{\label{fig:phasediagram}(слева) Магнитная фазовая диаграмма с нанесённой областью стабильности хопфионов с индексом Хопфа $\uHopf=2$; закрашенный кружок отмечает набор параметров хопфионов, показанных на рис.~\ref{fig:hopfion}; незакрашенный кружок показывает возможное начальное равновесное состояние, намагниченное в положительном направлении оси $OZ$, которое приложением отрицательного поля превращается в метастабильный однородный фон для хопфионов; на вставке --- зависимость параметров линий, очерчивающих область стабильности, от индекса Хопфа $\uHopf$ и асимптотики этих зависимостей.}
\end{figure}

Хопфионы энергетически выгодны, когда их энергия ниже энергии однородного фона. Соответствующая область в пространстве параметров $\uq$--$\uh$ ограничена двумя прямыми линиями
\begin{equation*}
 h^{-} = \frac{-q \upa-2 \mu ^2 \upms}{2 \upz},\quad
 h^{+} = h^{-} + \frac{\updm^2}{4\upex \upz}.
\end{equation*}
При этом расстояние между ними по вертикали равно $\Delta h = \updm^2/(4\upex \upz)$, а наклон $\ud \uh^{+}/\ud q = \ud \uh^{-}/\ud q = - \upa/(2\upz)$. Роль магнитостатического взаимодействия в этой простой модели сводится к смещению области стабильности вниз по вертикали на величину $\mu^2\upms/\upz$. Помимо этого, область стабильности хопфионов на однородном фоне ограничена условием стабильности самого фона $h>1-q$, соответствующим нестабильности относительно перехода в коническую фазу с последующим переходом в однородное состояние, намагниченное по полю. Это условие при увеличении $\mu$ смещает область стабильности хопфионов в область большей анизотропии $\uq$ и больших значений $|\uh|$. Поэтому в целом можно сказать, что эффект магнитостатического взаимодействия дестабилизирует хопфионы при заданных $\uq$ и $\uh$. Отметим, однако, что рассматриваемая простая модель основывается на пробной функции, не учитывающей магнитостатическое взаимодействие, и поэтому, вообще говоря, справедлива лишь при небольших $\mu^2$. Пример области стабильности для хопфионов с $\uHopf=2$ показан на рис.~\ref{fig:phasediagram}.

Отметим, что в работе~\cite{metlov2026tails} учитывалась лишь стабильность намагниченной против поля однородной фазы относительно возмущений с нулевым волновым вектором (однородного вращения), определяемая условием $h>-q$. В действительности, в гелимагнетике более энергетически выгодными оказываются возмущения с конечным волновым вектором, соответствующие переходу в коническую фазу. Это не меняет принципиальных выводов работы~\cite{metlov2026tails}, но сдвигает область стабильности хопфионов в область с бóльшими значениями $\uq$, как показано на рис.~\ref{fig:phasediagram}.

Там же показаны зависимости от индекса Хопфа и асимптотики трёх параметров, полностью определяющих область стабильности: $\Delta h$, $\ud h^{+}/\ud q$ и $h^{-}|_{q=0}$. В целом можно сказать, что с увеличением топологического индекса область стабильности хопфионов сжимается, поскольку $\Delta h$ стремится к нулю при $\uHopf\rightarrow\infty$. Однако убывает она так медленно ($\propto 1/\log^3(1+\uHopf)$), что фактически возможна реализация любых концентрических хопфионов со сколь угодно большим индексом Хопфа. У этой зависимости есть и интересная особенность: хопфионы с $\uHopf=2$ и $\uHopf=3$ имеют более широкую область стабильности, чем хопфионы с $\uHopf=1$. За счёт этого возможно ``просеивание'' хопфионов по топологическому индексу путём изменения величины магнитного поля.

Область стабильности также немного наклоняется при увеличении индекса Хопфа, и этот наклон при $\uHopf\rightarrow\infty$ выходит на конечный предел, а вот её смещение вниз при увеличении $\mu^2$ за счёт магнитостатического взаимодействия очень слабо зависит от $\uHopf$.

\section{Выводы}
В работе построена и решена простая аналитическая модель для нового типа магнитных хопфионов в гелимагнетике. Получены равновесные профили и построена область стабильности хопфионов в пространстве параметров: магнитное поле -- анизотропия с учётом магнитостатического взаимодействия.

Из проведённого анализа следует, что в гелимагнетике могут существовать концентрические хопфионы практически сколь угодно высоких порядков на фоне однородного состояния, намагниченного против приложенного магнитного поля. Также определён диапазон параметров материала, способного нести такие хопфионы. Это должен быть анизотропный гелимагнетик с низкой намагниченностью насыщения (поскольку магнитостатические эффекты дестабилизируют хопфионы) и относительно небольшой константой Дзялошинского–Мория.

Ни один из гелимагнетиков, широко исследуемых сегодня в поисках топологических состояний, этими свойствами не обладает, поэтому наблюдаемые в экспериментах хопфионы либо требуют дополнительной стабилизации в ограниченной геометрии, либо существуют на фоне неоднородных состояний намагниченности, что затрудняет их свободное перемещение. Более того, большинство материалов с требуемыми значениями анизотропии не покажут в равновесии типичных для гелимагнетика киральных фаз из-за недостаточной для этого силы взаимодействия Дзялошинского–Мория, а вот метастабильные хопфионы они поддерживать смогут.

Тем не менее, современное развитие материаловедения, в том числе и с применением нейросетевых технологий~\cite{Lyakhov2013}, даёт надежду на то, что материалы, удовлетворяющие этим критериям, могут быть найдены или сконструированы. Это, в свою очередь, откроет путь к трёхмерной спинтронике, в которой хопфионы, в силу своей неабелевости, могли бы стать ключевым элементом для реализации массивно-параллельных топологических вычислений.

Работа поддержана Российским научным фондом в рамках проекта РНФ 25-22-00076.

%